# Phenomenological theory of photodetection

M.V. Lebedev

Institute of Solid State Physics, Russian Academy of Sciences,

Chernogolovka, Moscow distr., 142432, Russia

The existing theory of photodetection is based on assumptions which look very natural, but are nevertheless not rigorously derived from the first principles. This theory works well in a lot of cases and is proven with a great number of experiments. The reasons to have doubt in the validity of this theory are nevertheless serious contradictions with some experiments and intrinsic problems of the theory, as it will be explained below. Modern theory of photodetection is based on the coherence of light. Coherence is a wave property, so one can state that modern theory of photodetection derives its main results from the wave nature of light but the particle nature of photons is added by the quantization of light waves in a finite box. The Einstein's theory of light absorption and emission by two-level atoms was, on the contrary, formulated exploiting the particle nature of light. The result of calculation of energy fluctuation of black body radiation in a closed volume found by Einstein showed that wave fluctuations indeed play a minor role at optical frequencies in comparison with particle fluctuations. According to modern theory light emission and absorption by the walls of a cavity kept at some temperature should be Poissonian, because these are results of light-matter interactions of a great number of independent light modes. A natural question arises: how can Poissonian light demonstrate Plank's law, which originates from correlations between photons in filling the phase space cells of the photon gas? Another strange result of modern photodetection theory is the independence of the detector quantum efficiency on the state of light detected. It will be shown that this comes from considering individual photons in a common quantum state of light as independent particles without determination of appropriate observables. Taking all this into account, one can state that there are some reasons to consider the existing theory as not complete and an attempt to improve it on the basis of the well known considerations of black body radiation given by Bose and Einstein is presented below.

**Modern theory of photodetection**

This theory consists of two independent parts. The first part: the calculation of the probability to detect one or more photons if the detector interacts with a given pure quantum state of the electromagnetic field during a very short time interval $\Delta t$. The second part: the calculation of a probability distribution of a number of photons detected during a finite time interval $T$ and a calculation of correlations between detection of one or more photons at different short time intervals $\Delta t_j$ if the field has thermodynamic fluctuations.

The first part is calculated in the following way [1]. At the first step a one-atom photodetector is considered and the probability $P\Delta t$ to detect a photon during a short time interval $\Delta t$ if a coherent state of the electromagnetic field with intensity $I$ interacts with this detector is calculated. This gives:

$$P\Delta t = \eta I \Delta t$$

where $\eta$ is the quantum efficiency of the detection process. At the second step an $N$-atom photodetector interacting with the coherent state of the field is considered and the binomial distribution for the number of detected photons is postulated.

$$P(n, \Delta t) = C_N^n (\eta I \Delta t)^n (1 - \eta I \Delta t)^{N-n}$$

Here $C_N^n$ are binomial coefficients. Supposing that $\eta I \Delta t \ll 1$ one gets:

$$P(n, \Delta t) \cong 0, \; n > 1; \; P(1, \Delta t) = N\eta I \Delta t$$

We see that only one photon detection is possible for infinitely small time intervals.

The second part of the photodetection theory exploits essentially the coherent states of electromagnetic field. If one considers correlations between measurements done at different small time intervals $\Delta t_j$ he should know the state of the field after every measurement, but the statement that any density operator of the field can be represented in coherent states, which are robust against photon absorption, reduces the problem to the calculation of the intensity correlation functions

$$G_n(t_1, \ldots t_n) = \langle I(t_1)I(t_2) \ldots I(t_n) \rangle$$

where the brackets denote the ensemble average.

The calculation for the probability $P(n, t, T)$ to detect n photons during a finite time interval $T$ leads to the Mandel's formula [1]:

$$P(n, t, T) = \int_0^\infty \frac{1}{n!} W^n e^{-W} \mathcal{P}(W) dW$$

Here

$$W = \eta \int_t^{t+T} I(t') dt'$$

and $\mathcal{P}(W)$ is the quasi-probability density for $W$.

**Contradictions with experiment**

Short after the famous experiments of Hanbury Brown and Twiss demonstrating the photon bunching effect Artem'ev found a two-photon peak in a response of a photomultiplier tube [2]. He explained the existence of this peak just as Hanbury Brown and Twiss did, as a result of a photon bunching effect. In his further works he studied some properties of this peak [3,4]. This peak was experimentally observed also in [5,6] were its dependence on the intensity and wavelength of the detected light was studied in detail. The problem is that this peak is definitely two-photon one but cannot be explained with Mandel's formula as a result of photon bunching, because it was observed at light levels insufficient for observation of photon bunching and also in experiments where spatial coherence of light was not guaranteed. It was also shown that its dependence on the light flux is linear at low intensities what contradicts the Mandel's formula.

**Some doubt in the validity of Mandel's postulates**

Mandel's formula is based on two main assumptions, namely:

1. The probability of photon detection is proportional to the intensity of light.
2. The probability to detect more than one photon during a sufficiently short time interval is at low intensity of light negligible.

These assumptions, strictly speaking, cannot be rigorously deduced from the fundamental physical principles. Moreover, the second assumption contradicts Artem'evs experimental finding. On the other hand, photon bunching predicted by Mandel's formula was experimentally proven in a lot of experiments. The only way to understand this situation is to suppose that both phenomena exist simultaneously but have different origin. One can say that Hanbury Brown and Twiss bunching is a wave bunching and demands spatial and temporal coherence but Artem'evs bunching is a particle one and can be observed without wave coherence.

**Bose definition of a phase space unitary sell is not equivalent to Mandel's definition of a coherence volume**

The definition of a unit phase space sell for the photon gas enclosed in a cavity given by Bose is quite different from the definition of "coherence volume" in modern theory of photodetection. The origin of the existence of such a unit sell, according to Bose, is the Heisenberg's uncertainty principle, hence the uncertainty has a spatial nature, because photons are enclosed in a finite volume. While spatial coherence in modern theory do depends on the dimensions of the radiating source and the distance between the source and the observation point, the coherence length along the light beam depends, in contrast, on the monochromaticity of radiation. If one considers a small hole in a cavity filled with thermal radiation as a light source the volume of elementary phase space sell according to Bose will be:

$$\Delta x \Delta y \Delta z \Delta p_x \Delta p_y \Delta p_z = h^3 \tag{1}$$

Where $\Delta x \Delta y \Delta z = V$ is the volume of the cavity, and $\Delta p_x, \Delta p_y, \Delta p_z$ – the uncertainties of the photon momentum. One can, of cause, suppose that $\Delta x$ and $\Delta y$ should be replaced by the appropriate dimensions of the hole, due to diffraction of light, but the definition of $\Delta z$ should coincide with that of Bose. The coherence volume in coherence theory, in contrast, is determined by the coherence length, which is the coherence time multiplied by the velocity of light. This coherence time depends not on the dimension of the cavity $\Delta z$, but on the properties of individual atoms, which emit radiation into the cavity.

Bose calculated the phase space volume for a photon gas enclosed in a volume $V$ as:

$$8\pi \frac{h^3 \nu^2}{c^3} V d\nu \tag{2}$$

Here $\nu$ is the frequency of light, $h$ - the Plank's constant and $c$ – the velocity of light. He found then the number of elementary cells belonging to the frequency interval $d\nu$ dividing this result by $h^3$. One can see that the frequency interval $d\nu$ determines the phase space volume but not the volume of the elementary cell in contrast to the modern coherence theory.

**A chaotic light source cannot demonstrate Plank's law**

Considering absorption of thermal radiation by the walls of the enclosing it cavity, one have to conclude that photon absorption events should be completely uncorrelated and the absorption process should be Poissonian, because one cannot consider the thermal field into the cavity as spatially coherent. Integration over time in the definition of the integral intensity of light entering the Mandel's formula is equivalent to the spatial integration over the surface of the cavity containing black body radiation due to ergodicity of thermal light. The latter is in fact the integration over the ensemble of realizations of optical field and should be equivalent to integration over time. This leads to the natural question: how can Poissonian absorption and emission give the Plank's law? Consideration of photons as independent particles leads to Wien's radiation law of black body radiation as was mentioned by Einstein [7], but not to the Plank's law.

**Photons appear sometimes as individual objects even in cases where they have no individual observables**

Suppose we have a beam splitter and a two-photon light state falling on it. If the photons have individual observables such as wave vectors or polarizations in this state one have to consider the probabilities of transmission and reflection for individual photons. But in the case where photons have no individual observables within the state it would be incorrect to consider the situation when one photon is transmitted whereas the second one reflected, because photons have no individual observables before the beam splitter but get such observables after it. The consequence of such treating of photons as independent entities with their own properties is that the quantum efficiency of the detector does not depend on the particular state of the light field detected.

**Why Mandel's theory should be considered as an approximation?**

Mandel's theory consists of two independent parts: the calculation of probabilities of photon registration under interaction of a given pure state of electromagnetic field with a detector and statistical averaging of this result over ensemble or time. The second part is confirmed in a great number of experiments demonstrating photon bunching in thermal radiation. The first part is not so good tested because most of existing photon detectors cannot definitely resolve simultaneous absorption of two or more photons. Artem'evs experimental finding belongs just to this first part. Statistical averaging based on the wave properties of particles seems to be correct in Mandel's theory, because the wave coherence successively works not only for photons but for pions [8] as well. This means that any other theory able to predict the probabilities of many-quantum absorption from a pure state should give the results of Mandel's theory if one don't distinguish between single photon and many-photon absorption and considers a many-quantum absorption event as an ordinary single photocount.

We can conclude that Mandel's theory is a good semiclassical approximation in which quantum effects are partly accounted for by axiomatic assumption that only single photons are absorbed during infinitesimally short time intervals and the probability of their absorption is governed with the intensity of light. Both this assumptions look quite probably but they are nevertheless postulated without derivation from fundamental physical principles. This approximation has intrinsic problems, but can account for some quantum effects in a simple way.

**Quantum or classical coherence?**

It should be mentioned that the answer to this question depends on the definitions of what is quantum and what classic and on what is considered: waves or particles. The wave picture based on Maxwell's equations is considered as classic for light and any effect which cannot be explained in this wave picture is considered as a quantum one. If we adopt this point of view, we say that the Hanbury Brown and Twiss effect is a classical wave coherence effect. The same effect considered for massive particles like pions, should be named quantum, because the classical particles picture does not involve interference. The same holds for a definition of a chaotic light source. If we suppose that individual atoms independently emit light waves we have to sum chaotic amplitudes and name this light source chaotic. But from the particles point of view the probability of emission of a particle is proportional to the square of the resulting amplitude and we immediately find that the emission of particles from a light source with a chaotic amplitude is correlated.

Our consideration shows that the coherence effects originating from the Bose-Einstein statistics of thermal radiation and from the coherence of light are different. The first ones originate from the uncertainty principle and the entropy maximum while the seconds from the trajectories indistinguishability. This difference can be illustrated with experiments with pseudothermal light [9] originating from a laser light scattered with a rotating ground glass disk. The stray light is obviously not a thermal one but do demonstrates a bunching effect. This chaotic light source demonstrates photon bunching without Bose – Einstein statistics, so we can conclude that this bunching effect originates from classical wave coherence. The important consequence of the above consideration is that both types of coherence can exist simultaneously.

**Formulation of the phenomenological theory of photodetection**

The problem of photodetection calculation from the first principles is rather complicated and a lot of simplification assumptions are usually made during the calculations. In my previously publications [10,11] an alternative way, based on a phenomenological point of view was developed. The resulting expressions are free from using the intensity of light and are able to explain the observations of the two-photon peak at extremely low light levels. It will be shown in what follows that they eliminate although the discrepancies between theory and experiment in absolute quantum photometry proposed by Klyshko and Penin [12].

Phenomenological theory of photodetection can be formulated on the basis of the well known considerations of thermal radiation of Bose and Einstein. Suppose we have a cavity the walls of which are kept at some constant temperature. Bose showed that one can get the Plank's law of black body radiation considering the phase space of thermal radiation filling the cavity considered as an ideal gas of photons, dividing this phase space in elementary cells of the volume of $h^3$ and calculating the filling of these cells consistent with the maximum of entropy. Photons are absorbed and emitted and the population of the cells thus obviously fluctuates in time. The probability $P_n$ that the phase space cell will be populated with $n$ photons is the known Bose-Einstein distribution:

$$P_n = \frac{1}{\langle n \rangle + 1} \left( \frac{\langle n \rangle}{\langle n \rangle + 1} \right)^n \tag{3}$$

The mean absorption rate of photons with given energy was found by Einstein while considering thermal equilibrium between thermal radiation and two-level atoms:

$$\langle R \rangle = \frac{\langle n \rangle}{\langle n \rangle + 1} \tag{4}$$

This value is connected with the number of excited atoms $N_e$ and the number $N_g$ of atoms in the ground state, if we suppose that the atom states are nondegenerate, as:

$$\frac{\langle n \rangle}{\langle n \rangle + 1} = \frac{N_e}{N_g} \tag{5}$$

Dividing the numerator and denominator of the right hand side of the above expression by the total number of atoms we get the following connection between probabilities of occupation of these states:

$$\frac{\langle n \rangle}{\langle n \rangle + 1} = \frac{p_e}{p_g} \tag{6}$$

This last expression shows that the mean transition probability is a conditional probability to find the atom in excited state provided it was initially in a ground state. It was found assuming a situation of uniform space filled with independent two-level atoms and electromagnetic field, which corresponds to a dilute ideal gas of radiating atoms. The situation changes if we consider thermal equilibrium between the electromagnetic field in a cavity and the walls of the cavity, because we have to account for reflection of light from the boundary of the cavity. Here we face a nontrivial problem: how does light cross the boundary between two media – as independent waves or as independent particles? Both answers are consistent with the Kirhhoff's law, but have quite different consequences [11]. We cannot neglect reflection simply because without reflection the cavity for light will not exist. To bypass this problem let us consider the following system. Suppose we have a cavity build of ideal reflecting mirrors. On the surfaces of the mirrors is situated a low density ensemble of two-level atoms, which can emit and absorb radiation. In this situation the phase space cells of the photon gas are well defined and we can consider the interaction between the atoms and field without considering the transmission and reflection of light at the boundary. We will account for these effects later, when considering light registration with real detectors.

In thermal equilibrium the mean absorption rate should be equal to the mean emission rate of photons. Expression (3) means that the probability $P_n$ to find a phase space cell populated with $n$ photons is the probability $P_0$ to find an empty cell multiplied by the $n$-th power of the mean probability of photon emission $\langle R \rangle^n$:

$$P_n = P_0 \langle R \rangle^n \tag{7}$$

Equations:

$$\sum_{n=0}^{\infty} P_n = 1 \tag{8}$$

$$\sum_{n=0}^{\infty} n P_n = \langle n \rangle \tag{9}$$

determine $P_0$ and $\langle R \rangle$ giving thus (3).

Emissions of individual photons are independent, provided the cell is found empty, because the mean conditional probability of $n$-photon emission is simply the product of $n$ one-photon mean emission probabilities:

$$\frac{P_n}{P_0} = \langle R \rangle^n \tag{10}$$

The structure of expression (3) shows that the mean transition rate does not depend on the phase space filling because

$$\frac{P_n}{P_{n-1}} = \langle R \rangle \tag{11}$$

does not depend on $n$. But the probability (3) to find a cell populated with $n$ photons definitely depends on its filling.

The above equations hold for thermal equilibrium between the two-level atoms at the walls of the cavity and the enclosed in the cavity electromagnetic field. These probabilities determine the mean number of equally populated phase space cells and belong to a set of statistical probabilities in a stationary system.

We can of course ask for another kind of probabilities. Suppose we have a cell populated with $n$ photons. What is the probability that k photons from this cell will be absorbed? This probability is a probability of a quantum mechanical interaction process between radiation and atoms and should obviously differ from (3). Actually, the theory of transitions in a quantum mechanical system affected by electromagnetic field contains a probability that no transition occurs despite of the existing of a field, but the set of probabilities (3) obviously contains no such a probability. This quantum mechanical probability should be a universal function of the phase space cell filling. The mean number of equally populated cells will be constant in thermal equilibrium. In other words the mean number of photons absorbed from a given phase space cell should be equal to the mean number of photons emitted into this cell.

The starting point of our consideration is the phenomenological assumption that the absorption probability of the ensemble of equally populated phase space cells depends on their filling $n$ and is simply:

$$R = \frac{n}{n+1} \tag{12}$$

We suppose in fact that the probability $W_k(n)$ of $k$-photon absorption from the ensemble of equally populated cells has the same structure as the probability (3) to find a cell populated with $n$ photons: this probability is the probability of no-absorption multiplied by the $k$-th power of the one photon absorption rate, while the mean number of absorbed photons is equal to $n$:

$$W_k(n) = \frac{1}{n+1}\left(\frac{n}{n+1}\right)^k \tag{13}$$

For $n = 1$ this gives:

$$W_k(1) = \left(\frac{1}{2}\right)^{k+1} \tag{14}$$

This means that a one-photon field state can produce $k$-quantum absorption with the probabilities above. It looks as contradicting the energy conservation law, but can nevertheless be true, because the probabilities we are discussing belong not to a closed quantum mechanical system but to a system interacting with a thermal bas. All probabilities considered here correspond to a measurement made within a small but finite time interval. The photons absorbed during this time interval can be absorbed at different time moments within this interval. We don't consider the time evolution of the field and atom states here. The only thing, we can discuss, are the statistical properties of light-matter interaction considered in a frame of statistical thermodynamics. In this sense we are seeking for the transition probabilities which determine the energy exchange between an ensemble of equally populated phase space cells and the atoms at the walls of the cavity. Expression (13) is the Bose-Einstein distribution again with the mean equal to the cell filling $n$. This value is not a small number in contrast to (3) and can explain photon bunching observed at rather small intensities of emitted light.

We can establish now the total probability of k-photon absorption over the whole number of phase space cells:

$$U_k = \sum_{n=0}^{\infty} P_n W_k(n) = \sum_{n=0}^{\infty} P_n \frac{1}{n+1}\left(\frac{n}{n+1}\right)^k \tag{15}$$

This is the probability of absorption of k photons from the thermal field by the atoms at the walls of the cavity normalized by the total number of phase space cells. This expression contains uncertainty for $n = k = 0$ but we can define in this case $\left(\frac{n}{n+1}\right)^k = 1$ and get a physically acceptable result:

$$U_0 = P_0 + \sum_{n=1}^{\infty} P_n \frac{1}{n+1} \tag{16}$$

The mean number of absorbed photons is:

$$\langle k \rangle = \sum_{k=0}^{\infty} k U_k = \sum_{k=0}^{\infty} k \sum_{n=0}^{\infty} P_n \frac{1}{n+1} \left(\frac{n}{n+1}\right)^k = \sum_{n=0}^{\infty} P_n \frac{1}{n+1} \sum_{k=0}^{\infty} k \left(\frac{n}{n+1}\right)^k =$$

$$\sum_{n=0}^{\infty} P_n \frac{1}{n+1} \frac{n}{n+1} \sum_{k=1}^{\infty} k \left(\frac{n}{n+1}\right)^{k-1} = \sum_{n=0}^{\infty} P_n n = \langle n \rangle \qquad (17)$$

Because

$$\sum_{k=1}^{\infty} k x^{k-1} = \frac{d}{dx} \sum_{k=1}^{\infty} x^k = \frac{d}{dx}\left(\frac{x}{1-x}\right) = \frac{1}{(1-x)^2} \qquad (18)$$

This coincides with the mean number of photons into the cavity.

It is interesting to clarify the difference in meaning between the probabilities $U_k$ and $P_n$. The letter are the probabilities to find a phase space cell of a photon gas into the cavity populated with $n$ photons, while the first ones are the probabilities to absorb $k$ photons from an equilibrium thermal field by the atoms at the walls of a cavity. $U_0$, for example, is the probability that no photons are absorbed. This probability should definitely exceed $P_0$, because photons may be not absorbed not only in the case when there are no photons in the cavity but also in the case when there are some but they are not absorbed. Actually:

$$U_0 = \langle \frac{1}{n+1} \rangle = -\frac{P_0}{\langle R \rangle} \ln(1 - \langle R \rangle) = \frac{P_0}{\langle R \rangle}\left(\langle R \rangle + \frac{\langle R \rangle^2}{2} + \cdots\right) = P_0 + \frac{1}{2}P_1 + \cdots \qquad (19)$$

Because:

$$\langle \frac{1}{n+1} \rangle = \sum_{n=0}^{\infty} \frac{P_n}{n+1} = P_0 \sum_{n=0}^{\infty} \frac{\langle R \rangle^n}{n+1} = \frac{P_0}{\langle R \rangle} \sum_{n=0}^{\infty} \frac{\langle R \rangle^{n+1}}{n+1} \qquad (20)$$

And:

$$\sum_{n=0}^{\infty} \frac{\langle R \rangle^{n+1}}{n+1} = \int \left(\frac{d}{d\langle R \rangle} \sum_{n=0}^{\infty} \frac{\langle R \rangle^{n+1}}{n+1}\right) d\langle R \rangle = \int \frac{d\langle R \rangle}{1-\langle R \rangle} = -\ln(1 - \langle R \rangle) \qquad (21)$$

We see that $U_0 \to P_0$ only if $\langle R \rangle \to 0$. The total absorption probability in this limit due to the completeness of the probability space will be $1 - U_0 = \langle R \rangle$. This coincides with (4), but the total absorption probability given by the set of probabilities $U_k$ should coincide with the mean absorption probability of the two-level atoms which are in thermal equilibrium with the radiation field at any arbitrary value of $\langle R \rangle$. To get this result we have to remember that $\langle R \rangle$ is a thermodynamic probability and its definition differs from the definition of the quantum mechanical probability in the following way. The thermodynamic probability cannot be zero if some electromagnetic field interacts with an ensemble of two-level atoms, while the quantum mechanical probability can. The thermodynamic probability can be determined as the mean number of absorbed photons during a short time interval, while the quantum mechanical probability is the probability to get a $k$-photon absorption event under interaction with a given pure state of the electromagnetic field. To get a thermodynamic probability of photon absorption we have to take the average over the quantum mechanical probabilities $W_k$. We obtain than that the thermodynamic probability to get $n$-photon absorption coincides with $P_n$. For $P_0$ the quantum mechanical probability coincides with the thermodynamic one. A natural way to eliminate the quantum mechanical probability is to consider an ensemble of equally populated phase space cells. The average of the absorption over this ensemble is just the thermodynamic probability.

For the one-photon absorption we get:

$$U_1 = \langle \frac{1}{n+1}\left(\frac{n}{n+1}\right)\rangle = \langle \frac{1}{n+1}\rangle - \langle \frac{1}{(n+1)^2}\rangle = U_0 - \frac{1}{\langle R \rangle}\int U_0 d\langle R\rangle = \frac{1}{4}P_1 + \cdots \tag{22}$$

Because:

$$\langle \frac{1}{(n+1)^2}\rangle = \frac{P_0}{\langle R\rangle}\sum_{n=0}^{\infty}\frac{\langle R\rangle^{n+1}}{(n+1)^2} = \frac{P_0}{\langle R\rangle}\int \sum_{n=0}^{\infty}\frac{\langle R\rangle^n}{n+1}d\langle R\rangle = \frac{1}{\langle R\rangle}\int U_0 d\langle R\rangle \tag{23}$$

We can conclude that the absorption probabilities $U_k$ does not coincide with the appropriate probabilities $P_k$ to find $k$ photons in a unit phase space cell. This looks natural, because the existing in the cell photons should not necessarily be absorbed during the momentary interaction with the cavity walls.

**A possible quantum state of the thermal field**

The linear dependence of the two-photon peak in the response of a photomultiplier tube on the intensity of light illuminating the photocathode [6] excludes its explanation trough the two-photon Fock state of the thermal field, because the probability $P_2$ of excitation of this state quadratically depends on intensity. An alternative way to explain it is to suppose that black body radiation consists of a mixture of field states

$$|Gn\rangle = \frac{1}{\sqrt{n+1}}\sum_{k=0}^{\infty}\left(\frac{n}{n+1}\right)^{k/2}|k\rangle \tag{24}$$

instead of a mixture of pure Fock states. This states can be called geometrical states, because their coefficients form a geometrical progression. What are the real states of the field in the cavity? This question seems to have answers depending on the basis used. If the states $|Gn\rangle$ form a complete basis (or an overcomplete one as coherent states do) one can express the thermal field into the cavity alternatively in a Fock basis or in a geometrical states basis. The letter one may turn out to be more convenient for consideration of photon absorption. These states should be a solution of the problem of interaction between the thermal field of the cavity and N atoms at the cavity walls. The value of $R = \frac{n}{n+1}$, which can be interpreted as a transition probability of this interaction, is much greater than the mean transition rate $\langle R \rangle = \frac{\langle n \rangle}{\langle n \rangle + 1}$. This means that the energy uncertainty during the interaction is significant. As a result two and more atoms can participate in this process giving rise to terms in (24) with $k > 1$.

**Application to a detector of photons**

If we have now a cold detector thermally isolated from the walls of the cavity and placed near a small hole into the cavity wall we can try to determine the probability to get a k-photon absorption event by this detector making some additional assumptions. The smallness of the hole makes sure that our detector does not disturb the thermal radiation field of the cavity. The detector is supposed to be cold enough and its own thermal radiation can be neglected. One can expect that the thermal light flux will be absorbed with our detector just like it is absorbed with the walls of the cavity but in contrast to the letter ones thermal emission from the detector to the cavity will be practically absent. We assume, that the main expression (15) still holds for the detector when modified in the following way. The phase space sell filling distribution $P_n$ should be replaced by some other distribution $\check{P}_n$ after the falling onto the detector radiation enters its active volume through its boundary and an additional coefficient η called quantum efficiency, should be introduced accounting for individual properties of the detector.

The question about the connection between $P_n$ and $\check{P}_n$ is nontrivial and depends on how light cross the boundary: as a particle flux or as a collection of waves? This question was considered in detail in [10]. A wave falling on a boundary generates two waves: a transmitted and a reflected one. If we denote the quantum state of falling light as $|L\rangle$, the transmitted wave will be $\sqrt{\tau}|L\rangle$ and the reflected $(\sqrt{1-\tau})|L\rangle$. If $|L\rangle$ is a one-photon state we can say that the transmitted wave is the probability amplitude wave for the transmitted particle and the reflected wave the one for the reflected particle. We can although say that the initial quantum state is transmitted with probability $\tau$ and reflected with the probability $1-\tau$. The wave and the particle pictures give thus the same result. This interpretation holds for a one-photon state, but for a many-photon state the wave and the particle pictures contradict each other. If we consider light as a flux of independent particles we have to expect that every particle can pass the boundary independent on the other. As a result some particles would be transmitted whereas the others reflected. So probability distributions to detect a given number of transmitted and reflected particles will arise in this case. In the wave picture, on the other hand, only two probability amplitudes hold: $\sqrt{\tau}|L\rangle$ for the transmitted wave and $(\sqrt{1-\tau})|L\rangle$ for the reflected one. Both the particle and the wave pictures agree with the Kirchhof"s law, but lead to different photon statistics. Bose-Einstein and Poissonian statistics does not change after reflection and transmission for a particles flux [13], whereas these are modified for waves [11]. In this last case we get:

$$\check{P}_n = \tau P_n \quad n \geq 1 \quad and \quad \check{P}_0 = (1-\tau) + \tau P_0 \tag{25}$$

Every light state corresponding to a unitary phase space cell is transmitted through the photocathode boundary with the probability $\tau$ and reflected with probability $1-\tau$. This results in:

$$U_k = \sum_{n=0}^{\infty} \tau P_n \frac{1}{n+1}\left(\frac{n}{n+1}\right)^k \quad k \geq 1 \tag{26}$$

It will be shown in what follows that the assumption that light cross the boundary as a wave leads to a possibility to explain the discrepancy between theory and experiment in measurements of the absolute quantum efficiency of photomultipliers described in [12].

Let us take now into account the photocathode characteristics. Suppose a one-photon state falls on the boundary of the photocathode and is transmitted with a probability $\tau$. It can give a photocount with a probability $\beta$. The total probability to get a photocount will be $\tau\beta$. The probability to get no photocounts will be:

$$1 - \tau + \tau(1 - \beta) = 1 - \tau\beta \tag{27}$$

The product $\tau\beta$ is usually called $\eta$, the quantum efficiency of the photocathode. This quantum efficiency should enter the resulting expression as the transmission probability $\tau$:

$$U_k = \sum_{n=0}^{\infty} \eta P_n \frac{1}{n+1} \left(\frac{n}{n+1}\right)^k \quad k \geq 1 \tag{28}$$

In fact we face here a fundamental problem: at what stage one should place the boundary between quantum and classical consideration of the detection process? Our last assumption that $\beta$ behaves as $\tau$ not for the detecting of a one-quantum state only, but for detecting of any quantum state is equivalent to the assumption that decoherence occurs somewhere after the photocathode of a photomultiplier [14].

The probability $\beta$ consists, according to our theory, of the fundamental part which depends on a quantum state of light detected multiplied by the probability to detect the photoelectron dependent on the efficiency of a given device. The fundamental probability consists of a series of probabilities to detect one, two, or more quanta. According to (13) the total fundamental probability to detect anything for $n$-quantum state is:

$$\sum_{k=1}^{\infty} W_k(n) = 1 - W_0(n) = \frac{n}{n+1} \tag{29}$$

For a one-quantum state this gives $\frac{1}{2}$. For the first glance this should mean that the upper limit for the quantum efficiency of a photocathode should be its fundamental quantum efficiency, namely $\frac{1}{2}$. This is not far from being true, because the free propagation length of a photoelectron in the material of the photocathode is of the order of 10 nm. That's why the thickness of semitransparent photocathodes is usually made of the same order. For reflection type photocathodes only a thin layer of material just after its boundary serves for photoelectron generation. All the photons absorbed in deeper layers of the photocathode give no photoelectrons. For devices with an internal photoeffect, such as avalanche photodiodes all photons absorbed can give a photocount and a fundamental limit for their quantum efficiency does not hold, because instead of the above consideration we have to integrate the device response over the whole light penetration length.

Here we can see also the difference between the quantum mechanical and the statistical probabilities. In absolute quantum photometry we measure the quantum mechanical probability, that is the probability to get a photocount as a result of interaction between the photocathode and a two- or one-quantum state of light. We measure the statistical probability while determining the quantum efficiency with a traditional method, because we simply count the mean number of photoevents during a long time interval, which is just the statistical probability. Suppose we illuminate the photocathode with weak thermal light. The mean number of photoelectrons will then be $\eta P_1$, while the mean number of photocounts $\frac{1}{2}\eta P_1$ and the mean number of one-electron pulses $\frac{1}{4}\eta P_1$. The experimental definition of quantum efficiency in a traditional method depends on how good can one discriminate one-electron pulses from many-electron ones. In the case of bad discrimination what is determined experimentally and called quantum efficiency is $\frac{1}{2}\eta$, while in the case of good discrimination of one-electron pulses this is $\frac{1}{4}\eta$. If we look into the data on quantum efficiency of different photocathodes we find that the best values actually lie between 20% and 40%.

The fundamental probability to detect one photoelectron from a one-photon state is, according to (14), $\frac{1}{4}$, so we can denote the one-quantum probability as $\frac{1}{4}\eta$. The fundamental probability to detect two photoelectrons from a one-photon state is $\frac{1}{8}$. This means that we have detected two quanta in sequence from a given phase space sell in a short time interval. One can ask of cause about the mechanism of repopulation of this given sell including its transition to the empty state after the absorption of the first quantum and subsequent population with one quantum again. The only thing which can be said about this is that the time interval between subsequent absorptions of photons should be short enough, because for long times we have to use for the probability of two-quantum absorption the product of the one-quantum absorption probabilities: $\frac{1}{4} \times \frac{1}{4} = \frac{1}{16}$ instead of $\frac{1}{8}$. This subsequent absorptions should be a result of some quantummechanical process, otherwise we have to use for the quantum efficiency $\eta^2$ instead of $\eta$. According to (26) the influence of the boundary will be the same as in the one-quantum case, but now two photoelectrons are generated and a question about the probability of their registration should be answered. In the case of independent registration one have to consider all possible probabilities for every photoelectron, whereas a common probability of registration holds if we consider a common quantum state of photoelectrons [14]. Experimental observations of the two-photon two-electron peak at rather low light levels [2-4,6] would be difficult to explain without the assumption that its probability is $\frac{1}{8}\eta$.

According to the modern theory of photodetection every photon is detected with a constant probability $\eta$ called quantum efficiency and independent on the state of light, so the probability to detect a one-photon state is $\eta$ and two one-quantum states $\eta^2$. This quantum efficiency is measured experimentally as a ratio of the average number of counts of the detector and the average power of the light beam illuminating the photodetector normalized by the quantum of energy of the detected photons $\hbar\omega$. These measurements are usually made at sufficiently low light levels when one-quantum states dominate in the light flux.

**Absolute quantum photometry: comparison between theory and experiment**

Klyshko and Penin used for absolute quantum photometry entangled photons – photon pares generated by a parametric down conversion process. The idea was rather simple: if you have two photons, generated simultaneously and propagating in known directions you can count them with two photomultipliers, acquire the number of counts of every photodetector and the number of coincidence counts as well. Denoting quantum efficiency as η we get

$$N_1 = \eta_1 M \quad \text{and} \quad N_2 = \eta_2 M \tag{30}$$

for the first and

$$N_c = \eta_1 \eta_2 M \tag{31}$$

for the second, where $M$ stands for the number of photon pares. Now we can find the quantum efficiencies of the photomultipliers as

$$\eta_1 = \frac{N_c}{N_2} \quad \text{and} \quad \eta_2 = \frac{N_c}{N_1} \tag{32}$$

This method was called by Penin and Klyshko a two-channel method, because every photon of the entangled pare was detected with a separate photomultiplier.

Beside this method they developed two one-channel methods where only one photomultiplier was used. In both one-channel methods the entangled pare was focused on the same photocathode. In this case sometimes one and sometimes two photoelectrons were emitted because $\eta \neq 1$. The probability to get one electron was:

$$p_{1e} = 2\eta(1-\eta) \tag{33}$$

while to get two:

$$p_{2e} = \eta^2 \tag{34}$$

Two-electron pulses at the output of a photomultiplier had twice the amplitude of the one-electron pulses, hence they could be counted separately from the latter. Dividing the number of two-electron pulses $N_{2e}$ by the number of one-electron ones $N_{1e}$ we get:

$$\frac{N_{2e}}{N_{1e}} = \frac{\eta}{2(1-\eta)} \tag{35}$$

and can now calculate the quantum efficiency $\eta$.

Another possibility is to count pulses without regarding their amplitudes but in one case (called case A) when both photons of the entangled pare fall on the photocathode:

$$N_A = 2\eta(1-\eta)M + \eta^2 M \tag{36}$$

And in the other case when one of the photons is blended by a mechanical chopper (case B):

$$N_B = \eta M \tag{37}$$

Then:

$$\frac{N_A}{N_B} = 2 - \eta \tag{38}$$

and we can again determine the quantum efficiency of the photomultiplier.

Klyshko and Penin measured with these methods a number of photomultipliers and compared their results with cathode quantum efficiencies of the tubes given in their technical characteristics and measured with traditional method. The result of this comparison is shown in Table 1, taken from [12]. One can see a systematic deviation of quantum efficiencies measured with the methods of Penin and Klyshko from the data of conventional measurements. Penin and Klyshko attributed this deviation to the quantum efficiency lowering due to degradation of photomultipliers with time, because all tested photomultipliers were rather old (10 years).

Table 1.

| PMT type | Integral Radiant Sensitivity (mA/lm) | Quantum Efficiency, PMT Sheet (%) | Quantum Efficiency, one-channel method (%) | Quantum Efficiency, two-channel method (%) |
|---|---|---|---|---|
| PMT-79 | 0,31 | 7,8 | 3,6 | 3,8 |
| PMT-79 | 0,23 | 6,0 | 3,3 | 3,0 |
| PMT-79 | 0,20 | 5,3 | 1,8 | - |
| C31034 A (USA) | 0,7 | 18,0 | 7,0 | 7,5 |

It would be interesting to compare their results with the phenomenological theory of photodetection outlined above. The entangled photons belong to a common quantum state which is in fact a superposition of two coupled waves. Every wave is transmitted through the boundary with a transmission coefficient $\tau$. The probability of a two-photon absorption will be thus proportional to $\tau^2$, whereas a two-quantum absorption from a single wave state is impossible because in contrast to detection of thermal light detecting of a biphoton is detecting of a pure quantum state and energy conservation holds for the detecting process. In the case of detecting thermal light two-quantum absorption from a single wave state is possible and proportional to $\tau$. This explains the unexpected experimental finding that the counting rate of two-photon two-electron pulses is much higher when registering thermal light in comparison with registration of entangled photon pares from a parametrical down conversion process [6]. If both entangled photons are falling on the same photocathode the probability to get only one photoelectron will be:

$$p_{1e} = 2\eta \frac{1}{4}(1-\eta) \qquad (39)$$

The probability to get two photoelectrons each from one of the entangled photons will be:

$$p_{2e} = \frac{1}{8}\eta^2 \qquad (40)$$

Thus we have:

$$\frac{p_{2e}}{p_{1e}} = \frac{\eta}{4(1-\eta)} \qquad (41)$$

Comparing (41) and (35) we find that for small quantum efficiencies (41) gives approximately twice the value calculated from (35).

For the two-channel method we have:

$$N_1 = \frac{1}{4}\eta_1(1-\eta_2)M \qquad (42)$$

$$N_2 = \frac{1}{4}\eta_2(1-\eta_1)M \qquad (43)$$

$$N_c = \frac{1}{8}\eta_1\eta_2 M. \qquad (44)$$

Now

$$2\frac{N_c}{N_2} = \frac{\eta_1}{1-\eta_1} \text{ and } 2\frac{N_c}{N_1} = \frac{\eta_2}{1-\eta_2} \tag{45}$$

This is in agreement with the result of the consideration of the one channel method above.

It is not so clear how should one analyze the second one-channel method of [12]. We can describe the interaction of the entangled two-photon state with the photocathode, but how can we describe the influence of the shutter that blends one of the photons of the entangled pare on our measurement? In this case we can try to guess the answer in the following way. Suppose we blend the photon not with a shutter but with a photocathode of another photomultiplier with quantum efficiency $\eta_x$. Then we can write:

$$N_A = 2\frac{1}{4}\eta(1-\eta)M + \frac{1}{8}\eta^2 M \tag{46}$$

And

$$N_B = \frac{1}{4}\eta(1-\eta_x)M + \frac{1}{8}\eta\eta_x M \tag{47}$$

Dividing (46) by (47) we get:

$$\frac{N_A}{N_B} = \frac{2-\frac{3}{2}\eta}{1-\frac{1}{2}\eta_x} \tag{48}$$

We see that the result depends on $\eta_x$, the quantum efficiency of the additional photomultiplier. We get substantially different results for $\eta_x = 0$, what can be interpreted as a 100% reflecting mirror, and for $\eta_x = 1$ which corresponds to the case of ideal absorber. Entangled photons form a common quantum state and the blended photon influence the result of the measurement in spite of the fact that it is not registered with the main photomultiplier.

We can try to guess: what kind of blending is necessary to get from the discussed one-channel method the same quantum efficiency of the photocathode as from the other two methods. For this purpose let us denote the quantum efficiency entering (32), (35) and (38) as $\eta_{old}$ and entering (41), (45) and (48) as $\eta_{new}$. Excluding the experimental ratios of different types of measurements, which are the same for the old and the new expressions for the quantum efficiency, we can find the relation between $\eta_{old}$ and $\eta_{new}$. For the two-channel method as well as for the first of the one-channel methods we get the same result:

$$\boldsymbol{\eta_{new}} = \frac{2\eta_{old}}{1+\eta_{old}} \tag{49}$$

We can now postulate that this expression holds for the second one-channel method also. This gives:

$$\boldsymbol{\eta_{new}} = \frac{2\left(2-\frac{N_A}{N_B}\right)}{1+\left(2-\frac{N_A}{N_B}\right)} \tag{50}$$

And we get:

$$\frac{N_A}{N_B} = \frac{2-\frac{3}{2}\eta_{new}}{1-\frac{1}{2}\eta_{new}} \tag{51}$$

Comparing (48) and (51) we conclude that $\eta_x = \eta_{new}$. Now we can reconstruct from (47) the expression for $N_B$:

$$N_B = \frac{1}{4}\eta(1-\eta)M + \frac{1}{8}\eta^2 M \tag{52}$$

Expression (52) gives essentially the number of one-electron pulses of the photomultiplier, because the second term accounts now for correlation between different detectors. But this number depends on what has happened with the blended photon, despite the fact that the number of photons registered with the additional photomultiplier is not measured – an interesting and unexpected result which can be in principle proven experimentally.

Finally we summarize the results of a new definition of quantum efficiency in comparison with its PMT datasheets and old measured values in Table 2.

Table 2.

| PMT type | Quantum Efficiency, one-channel method (%) | Quantum Efficiency, two-channel method (%) | Quantum Efficiency, PMT Sheet (%) | $\eta_{new} = \frac{2\eta_{old}}{1+\eta_{old}}$ for the one-channel method (%) |
|---|---|---|---|---|
| PMT-79 | 3,6 | 3,8 | 7,8 | 7,0 |
| PMT-79 | 3,3 | 3,0 | 6,0 | 6,4 |
| PMT-79 | 1,8 | - | 5,3 | 3,5 |
| C31034 A (USA) | 7,0 | 7,5 | 18,0 | 13,1 |

One can state that the systematical deviation of the datasheet values of the quantum efficiencies of the photocathodes from the measured with the method of absolute quantum photometry disappears if we calculate the quantum efficiency with the phenomenological theory of photodetection.

In subsequent experiments on absolute measurements of photomultiplier quantum efficiency [15] the experimental schema was radically changed. None of the one-channel methods was used and in the two-channel method one of the photomultipliers was replaced with a silicon APD (avalanche photodiode). As a result, the discrepancy between theory and experiment was eliminated and the measured values of the quantum efficiency coincide with values measured with conventional methods. This fact has a simple explanation. Photocathodes of the photomultiplier tubes have a very thin working layer (typically 10 nm), so that semitransparent photocathodes are very similar to atoms at the walls of the reflecting cavity – the model which was considered above. This means that two photons have to coincide at the photocathodes within a very short time interval to give a correlated photoemission of two electrons with the probability $p_c = \frac{1}{8}\eta_1\eta_2$. Otherwise the probability to count two photons will result from two uncorrelated one-photon detections with the probability $p_{2uncorrelated} = \frac{1}{16}\eta^2 = \frac{1}{4}\eta \times \frac{1}{4}\eta$. This probability is practically absent if one use two photomultipliers, but is significant in the case of an APD, because the avalanche triggered by a photon can occur not in a thin layer only but in a whole volume of the APD's p-n junction which has the thickness of the order of 1 micron. If uncorrelated two-photon detection dominates one gets (30) - (31) instead of (42) – (44) (factor $\frac{1}{4}$ simply enters then the definition of the quantum efficiency).

**Conclusions**

In conclusion, this work presents an attempt to give a phenomenological theory of photon counting by a photomultiplier free from using the known theory of optical coherence and the Mandel's formula. This does not mean that the results of the conventional theory are incorrect, because the letter ones hold for the case when the detector is placed far enough from the radiating source, where the coherence properties of the field do play a significant role. In our case when the detector is placed just in the vicinity of the radiating cavity the coherence effects are negligible and the main role plays the particle nature of light. The possibility of the presented theory to explain the experimental observations of a two-photon peak at extremely low light levels and to eliminate the discrepancy between theory and experiment in a method of absolute quantum measurement of quantum efficiency of photomultipliers shows that the particle nature of light should be taken into account as well as the wave one.